\documentclass[12pt]{article}
\usepackage{amsmath,amssymb,amsfonts,epsfig,graphicx}
\newcommand{\be}{\begin{equation}}
\newcommand{\ee}{\end{equation}}
\newcommand{\fixme}[1]{\textbf{FIXME: }$\langle$\textit{#1}$\rangle$}

\newcommand{\bse}{\begin{subequations}}
\newcommand{\ese}{\end{subequations}}
\newcommand{\bea}{\begin{eqnarray}}
\newcommand{\eea}{\end{eqnarray}}
\newcommand{\ba}{\begin{array}}
\newcommand{\ea}{\end{array}}

\def\la{\lambda}

\def\half{\frac{1}{2}}
\def\cN{{\cal N}}
\def\Ne{${\cal N}=8$\ }
\def\3dNe{$3d,\ {\cal N}=8\ $}

\def\sun2{$su(N)\times su(N)$}
\def\un2{$u(N)\times u(N)$}

\def\cs{Chern-Simons }

\def\id{1\!\!1}
 \makeatletter \@addtoreset{equation}{section}

\makeatother
\begin{document}
\begin{titlepage}
\vspace*{1mm}%
\hfill%
\vbox{
    \halign{#\hfil \cr
\;\;\;\;\;\;\;\;\;\; IPM/P-2009/012 \cr \;\;\;\;\;\;\;\;\;\;
NSF-KITP-09-50\cr arXiv:0904.4605 {\tt [hep-th]} \cr
           \cr
           } 
      }  
\vspace*{10mm}%
\begin{center}
\textbf{\textsl{{\Large On Half-BPS States of the ABJM Theory}}}
\end{center}
\vspace*{7mm}
\begin{center}
{\bf \large{M. M. Sheikh-Jabbari$^{1}$ and Joan Sim\'on$^{2,3}$ }}
\end{center}
\begin{center}
\vspace*{0.4cm} {\it {$^1$School of Physics, Institute for Research
in Fundamental Sciences (IPM)\\
P.O.Box 19395-5531, Tehran, IRAN\\
$^2$ School of Mathematics and Maxwell Institute for Mathematical Sciences,\\
King's Buildings, Edinburgh EH9 3JZ, Scotland\\
$^3$ Kavli Institute for Theoretical Physics, \\
University of California, Santa Barbara CA 93106-4030, USA}}\\
{E-mails: {\tt jabbari@theory.ipm.ac.ir, j.simon@ed.ac.uk}}%
\vspace*{1.5cm}
\end{center}

\begin{center}{\bf Abstract}\end{center}
\begin{quote}
We analyze $SU(2)$ invariant half-BPS states of the \3dNe or $\cN=6$ SCFT within  the
radial quantization of the ABJM theory \cite{ABJM}, the theory
proposed to describe $N$ M2-branes in the $R^3\times
\mathbb{C}^4/Z_k$ background. After studying the classical moduli space of these configurations,
we explicitly construct a set of gauge invariant operators involving 't Hooft monopole operators corresponding to these states. We show there is a one--to--one correspondence between the two sets
carrying R-charge $J$ and that they are labeled by Young tableaux of $J$ boxes with a maximum of $N$ rows. Restricting the full path integral to this half-BPS sector of the theory, we show
the latter is described in terms of $N$ fermions in a $2d$ harmonic potential in the
sector of vanishing angular momentum. The same classification, though in the $N\to\infty$
limit, arise from the plane-wave (BMN) Matrix theory as well as  the
11 dimensional LLM bubbling geometries \cite{LLM}, providing
supportive evidence for the ABJM theory and/or the Matrix model.

\end{quote}
\end{titlepage}\baselineskip 18pt%

\tableofcontents

\section{Introduction}

The \Ne3d SCFT is the theory describing  the low energy limit  of
multiple M2-branes in $R^{1,10}$. Moreover, it is expected to be the conformal
field theory dual to M-theory on the $AdS_4\times S^7$ background.
As such the recent proposals for a non-trivial interacting \Ne3d
field theory \cite{BLG}, the BLG theory, has prompted an extensive
study of these models. The original BLG theory, with totally
antisymmetric three-bracket structure \cite{BLG} and a three-algebra
with positive definite metric only describes dynamics of two
M2-branes \cite{3algebra-nogo}. The way for constructing a theory
which describes a generic number of M2-branes was paved by the
realization that the BLG theory may also be written as a
$SU(2)\times SU(2)$ $3d$ (supersymmetric) Chern-Simons gauge theory
with the $SU(2)$'s to have levels $k$ and $-k$ \cite{Mark}.

The generalization to $N$  M2-branes, for which the natural guess
would be an $SU(N)\times SU(N)$ supersymmetric Chern-Simons gauge
theory, now known as the ABJM theory, was proposed in \cite{ABJM}
shortly after the BLG theory. It was shown, through a construction
involving $N$ D3-brane intersecting an NS5-brane and a $(1,k)$
5-brane of type IIB theory and uplifting the system to M-theory and
taking the low-energy limit, that the theory describing $N$
M2-branes probing a (supersymmetric) $C^4/\mathbb{Z}_k$ orbifold is
a supersymmetric $U(N)_k\times U(N)_{-k}$ Chern-Simons theory
\cite{ABJM}. This theory, which is closely resembling the BLG theory
for the $N=2$, has $\cN=6$ supersymmetry, it is a conformal field
theory and it is invariant under the $Osp(4^*|6)$ superalgebra
\cite{Schwarz-SUSY}. For $k=1,2$ the ABJM theory is expected   to
become an \Ne3d theory. This theory has passed many tests and
many extensions of the model (to less supersymmetric Chern-Simons
gauge theories) have also been studied.

In this note we study and classify all the half-BPS configurations
of the ABJM theory which are invariant under the $SU(2|3)$ superalgebra
and compare it with the known results from the plane-wave matrix
theory \cite{BMN} and the half-BPS deformations of the eleven
dimensional plane-wave background \cite{LLM}. We show that there is
a one--to--one correspondence between these three. The half-BPS states
of an \Ne3d theory are labeled by the only quantum number they carry
$J$, which is the R-charge corresponding to a $U(1)\in SO(8)$
R-symmetry group of the theory. Being BPS the scaling dimension of
these operators $\Delta=J/2$  is protected by supersymmetry.

To study the half-BPS states of the ABJM theory we need to consider
monopole (or 't Hooft loop) operators. In the radial quantization of
the ABJM theory (i.e. the ABJM theory on $R\times S^2$) these are
operators which involve a non-zero magnetic flux on the $S^2$. Due
to the presence of the 't Hooft loop operator, these half-BPS
operators may seem to be non-local. However, since we are dealing
with a Chern-Simons theory their non-local part is a gauge artifact
\cite{ABJM, extra}.\footnote{For recent work on the construction of gauge invariant operators describing warped M2-branes see \cite{recent}.}

As we will show the half-BPS states with R-charge $J$ are
constructed from monopole operators the total magnetic flux of which
over the $S^2$ is $J$. Here we will give a detailed construction of
half-BPS operators and their classification by all possibilities
available for monopole operators of flux $J$. As we will discuss
such monopole operators are classified by the partition of $J$ into
$N$ non-negative integers ($N$ is the rank of the gauge group in the
corresponding ABJM theory). Therefore all the half-BPS states of the
ABJM theory, for any $k$, are labeled by Young tableaux of $J$
number of boxes and maximum $N$ number of rows.

It is well established that (e.g. see \cite{LLM,LLMvsLLL}) the
half-BPS sector of the $\cN=4$ $SU(N)$ SYM is equivalent to a system
of $N$ $2d$ fermions in the  Lowest Landau Level. In this work we
show that a similar  $2d$ fermionic picture is also true for the
half-BPS sector of the ABJM theory (for any $k$). In this case, in
contrast to the $\cN=4$ SYM case, the fermions are bound to move in
a $2d$ harmonic oscillator potential in the sector with zero angular
momentum.

The ABJM theory at level $k=1,2$ is a theory dual to M-theory on
$AdS_4\times S^7$ (or its Penrose or plane-wave limit, the $11d$
plane-wave) and as such one  expects to have a similar
classification for half-BPS states of the latter  theory. Although
an independent formulation for M-theory on $AdS_4\times S^7$ is
still lacking, for this purpose one can use the plane-wave matrix theory
as the discrete light cone quantization (DLCQ) of M-theory on the
$11d$ plane-wave background. \footnote{The $11d$ plane-wave and the
$AdS_4\times S^7$ are related by taking the Penrose limit. The
process of taking the Penrose limit closely resembles that of going
to an infinite momentum frame and/or the DLCQ \cite{TGMT}. The
plane-wave matrix model can also be considered as the DLCQ of
M-theory on the $AdS_4\times S^7$.} The half-BPS states of the
latter have been classified and shown to be all labeled by the
$J\times J$ representation of $SU(2)$ \cite{DSV-1}. These
representations are labeled by Young tableaux of $J$ boxes. In
contrast to the ABJM case, there is no restriction on the number of
rows of the Young diagrams in this case.

As the third description for these half-BPS states, we consider the
class of $11d$ supergravity solutions which are half-BPS
deformations of $11d$ maximally supersymmetric plane-wave. These
geometries are the $11d$ LLM (bubbling) geometries \cite{LLM}. As
discussed in \cite{LM} (see also \cite{Mark-M5,mark-coarsegraining}) these geometries
are labeled by a set of integers specifying the number of spherical M2
or M5-branes in the background. We show how this information can
naturally be encoded in a Young tableau, in perfect agreement with
the previous two descriptions.

This paper is organized as follows. In section 2, after reviewing
the ABJM theory we focus on its half-BPS sector and construct all
the half-BPS operators in the sector with R-charge $J$. In section
3, we show that the half-BPS sector of the ABJM theory is described
by a $2d$ fermion system. In section 4, we review the results of the
plane-wave matrix theory and its half-BPS states. Moreover, we
review the $11d$ LLM bubbling geometries and show that the half-BPS
deformations of the $11d$ plane-wave can be described by a Young
tableau, similarly to the one used for half-BPS states of the ABJM
theory. In this way we give a natural interpretation for the
monopole operators of the ABJM theory in terms of spherical M2 or M5
-brane giant gravitons. The last section is devoted to discussions.

\section{Half-BPS sector in the ABJM theory}

The ABJM theory is a supersymmetric $3d$ \cs theory with four
complex scalars $Z^A$ and four two component real $3d$ fermions $\psi_A$ in the
${\bf 4}$ of the $SO(6)$ R-symmetry group. These matter fields transform in the
bi-fundamental  $(N, \bar{N})$ representation of
$u(N)\times u(N)$ or its complex conjugate $(\bar{ N}, N)$.
Besides these propagating fields, there are a couple of non-dynamical
\cs gauge fields $A^{(1)}$ and $A^{(2)}$ in the $N\times N$ representations of each of
the $u(N)$ algebras. They have a Chern-Simons action with opposite integer levels for the
two gauge groups, $k$ and $-k$.

These theories are dual to $AdS_4\times S^7/Z_k$. For the particular values of $k=1,2$, the R-symmetry group is enhanced to $SO(8)$ and the number of supersymmetries to 32.

We are interested in studying the half-BPS sector of these theories
preserving $SO(3)\times SU(4)$, for $k=1,2$ or $SO(3)\times SU(3)$
for $k\neq 1,2$. These are states saturating the BPS bound $\Delta =
J/2$, where $J$ corresponds to their $U(1)$ R-charge and $\Delta$ to
their conformal dimension. Since R-charge acts as rotation on the
complex scalar fields, the preserved symmetries guarantee that such
states will only involve a single scalar field $Z=Z^1$, carrying one
unit of R-charge and having conformal dimension $1/2$.

It is convenient to use radial quantization by defining these
theories on $R\times S^2$ so that there exists an operator--state
correspondence. In addition, the eigenvalues of the Hamiltonian
correspond to the scaling dimensions of the operators. The classical
action will involve an extra mass coupling of the scalar fields to
the curvature of the 2-sphere \cite{ABJM,Ahmad,Berenstein:2008dc}.
The bosonic truncation of the total action to this
single complex scalar field $Z$ coupled to the two gauge fields is%
\be\label{Z-action}%
 S=-\frac{1}{8\pi}\int\ dt d^2\Omega\,
Tr\left[D_\alpha Z D^\alpha {\bar Z}+D_\alpha {\bar Z} D^\alpha Z -
\frac14({\bar Z} Z+Z{\bar
Z})\right]-S_{CS}\ , %
\ee%
 where $S_{CS}$ stands for the Chern-Simons piece%
\be%
S_{CS}=\frac{k}{8\pi}\,\int dt d^2\Omega\ Tr\sum_{i=1}^2
(-1)^{i+1}\,\left(A^{(i)}\wedge
dA^{(i)}+\frac23 A^{(i)}\wedge A^{(i)}\wedge A^{(i)}\right), %
\ee%
and the covariant derivatives are defined according to the field representation,
\be\begin{split}%
D_{\alpha} Z&=\partial_\alpha Z+i A^{(1)}_\alpha Z-iZ A^{(2)}_\alpha
\ ,\cr D_{\alpha} {\bar Z}&=\partial_\alpha {\bar Z}+i
A^{(2)}_\alpha {\bar Z}-i{\bar Z} A^{(1)}_\alpha\ .\
\end{split}\ee%
We have chosen the radius of the $S^2$ such that the mass is
(formally) equal to one half.

\subsection{Classical moduli space of half-BPS configurations}

In the following, we will solve the classical equations of motion
derived from \eqref{Z-action} while preserving the appropriate
bosonic symmetries. We will then check that these configurations
preserve half of the supersymmetries.\footnote{Our analysis is close
in nature to the one presented in Section III of
\cite{Berenstein:2008dc}. Here, however, we directly focus on the
subset of degrees of freedom relevant for the description of
half-BPS states with the appropriate bosonic symmetries.}

Classical half-BPS configurations must be $SU(2)$ invariant. As
such, the matrix $Z$ must be covariantly constant on the 2-sphere,
i.e. $D_\theta Z = D_\phi Z = 0$. Non-vanishing R-charge requires a
non-trivial time dependence on $Z$ to describe the rotation in the
$\{Z,\,\bar{Z}\}$ subspace. Since \eqref{Z-action} contains a \cs
term, gauge fields cannot be turned off consistently.
Their equations of motion are %
\be\label{F-J-eom}%
\begin{split}%
\epsilon^{\mu\nu\alpha}F^{(1)}_{\mu\nu}
&=\frac{2\pi i}{k} J^{\alpha}=\frac{2\pi i}{ k}\left({\bar Z}
D^{\alpha}Z-(D^\alpha {\bar Z}) Z\right)\ ,\cr
\epsilon^{\mu\nu\alpha}F^{(2)}_{\mu\nu} &=-\frac{2\pi i}{k}{\bar
J}^{\alpha}=-\frac{2\pi i}{k}\left({Z} D^{\alpha}{\bar Z}-(D^\alpha
{ Z}) {\bar Z}\right) .
\end{split}\ee%
Notice that $D_\theta Z= D_\phi Z = 0$ is on-shell equivalent to the
absence of covariantly constant vector fields $F_{t\mu}^{(i)}$
on the 2-sphere. Since $J^{\alpha}$ is a
conserved current, we learn that the $N\times N$ matrices%
\be\label{flux-matrix}%
F^{(i)}\equiv \int_{S^2} F^{(i)}_{\theta\phi}%
\ee%
are constants of motion, that is $D_t F^{(i)}=0$. Note that on a
2-sphere we have an invariant two-form, its volume-form, and hence
$F_{\theta\phi}\propto \sin\theta$ is not ruled out by the $SO(3)$
invariance condition.\footnote{This should be contrasted with the
half-BPS sector of $\cN=4$ SYM on $R\times S^3$, in the sense that
there are no invariant two-forms on the $S^3$.\label{footnote2}}

Let us next consider the equations of motion for the $Z$ field which
in the absence of the gauge fields take the form
$$
-\partial_t^2 Z+\nabla^2 Z-\frac14 Z=0 .%
$$
One may use the \un2 gauge transformations to diagonalize  $Z$. Let
us work in a gauge where
\be\label{Z-diagonal}%
Z=diag(z_1,z_2,\cdots, z_N),\ \ z_i=e^{-i\omega_i t} w_i\  .%
\ee%
we learn that the spectrum of $Z$ is
$$
\omega_l=\sqrt{l(l+1)+\frac14}=l+\frac12\ .
$$
Hence for the half-BPS sector, where all the fields should be
constant on the $S^2$ (i.e. $l=0$) $\omega_i$ are all equal to
$\frac12$ in units of the radius of the $S^2$.

Working in the gauge $A^{(1)}=A^{(2)}$, conditions $D_\theta
Z=D_\phi Z=0$ are manifestly satisfied for $Z$'s in the half-BPS
sector. This gauge is preserved by a diagonal $u(N)$, which can be
used to set $A_0^{(i)}=0$ (this justifies the validity of the scalar field equation used above).
Thus $D_0 Z = \dot{Z}$. For our diagonal
configurations, we have
\be\begin{split}%
F^{(1)} &=\frac{\pi i}{k}\left({\bar Z} \dot Z-{\dot{\bar Z}}
Z\right)\cr F^{(2)}&=-\frac{\pi i}{k}\left({Z} {\dot{\bar Z}}-{\dot{
Z}} {\bar Z}\right)\,,
\end{split}\ee%
from which we conclude%
\be\label{flux-z2}%
 F^{(1)}=F^{(2)}=\frac{2\pi}{k}
diag(|z_1|^2,|z_2|^2,\cdots, |z_N|^2)\equiv 2\pi\
diag(n_1,n_2,\cdots, n_N),\qquad n_i\in\mathbb{Z}_+\ , %
\ee%
where quantization of the gauge field flux over the 2-sphere was
used in the last step, implying that%
\be%
|z_i|^2=k n_i\ .
\ee%
That is, $|z_i|^2$ is an (integer) multiple of the level $k$. From
the above equation we learn that the quantized fluxes of the gauge
fields $n_i$ are a collection of \emph{non-negative integers}.
Acting with the permutation group $S_N$, we can arrange them such
that $n_1\geq n_2 \geq \dots \geq n_N$. We would like to stress
that, being the eigenvalues of the matrix $F$, $n_i$ are all gauge
invariant quantities.

For the diagonal configurations specified by the set of fluxes
$\{n_i\}$ the
energy of the system is given by%
\be\label{Hamiltonian-ni}%
H=\frac{1}{16\pi} \int_{S^2} d^2\Omega\ Tr({\bar Z} Z+Z{\bar Z})\
= \frac{k}{2} \sum_{i=1}^N n_i\ . %
\ee%
Note that the energy is also related to the total flux of the gauge
fields over the two sphere,  $H= \frac{k}{8\pi} \left(Tr F^{(1)}+
TrF^{(2)}\right)$. One may also work out the angular momentum
associated to the $u(1)$ rotation of the $Z$%
\be\label{ang-mom}%
J=\frac{1}{4\pi}\int d^2\Omega\ Tr(\bar ZD_0Z-\overline{D_0Z}
Z)=\sum_{i=1}^N |z_i|^2=k\sum_{i=1}^N n_i\ . %
\ee%
It is readily seen that $H=J/2$, which is the BPS condition.

\paragraph{Supersymmetry: } So far we have argued that being in the
half-BPS sector demands turning on an $SO(3)$ invariant mode of only
one of the four complex scalars $Z^A$. Here we show that this is
indeed enough for being half-BPS. To see this consider the
supersymmetry variations for the fields in ABJM theories defined on
$R\times S^2$ written in \cite{Ahmad}. The amount of
supersymmetry preserved by any bosonic configuration is obtained by
solving
\begin{equation*}
\delta\psi_{Bd} = \gamma^\mu D_\mu Z^A_d\,\epsilon_{AB} +
f^{ab\bar{c}}{}_d\,Z^C_a\,Z^A_b\,\bar{Z}_{C\bar{c}}\,\epsilon_{AB} +
f^{ab\bar{c}}{}_d\,Z^C_a\,Z^D_b\,\bar{Z}_{B\bar{c}}\,\epsilon_{CD} -
\frac{1}{3}Z^A_d\,\gamma^\mu\nabla_\mu\epsilon_{AB}=0\ .
\end{equation*}
In the above $a,b,\cdots$  are denoting the $u(N)\times u(N)$
bi-fundamental indices, $f^{ab{\bar c}}{}_d$ are the structure
constants built from the $u(N)$ symmetric traceless and totally
anti-symmetric three tensors \cite{BL3} and $\epsilon_{AB}$ are
fermions on $R\times S^2$ as well as being in the ${\bf 6}$ of
$SU(4)$ R-symmetry (denoted by $A,B$ indices). In particular note
that $\nabla_\mu\epsilon = \gamma_\mu\,\epsilon/2$, and
$\gamma_0\epsilon=i\epsilon$ and therefore, there are 12 independent
$\epsilon$'s. Since
there is only one scalar $Z^1=Z$ turned on the above reduces to%
\be\label{SUSY-variation}%
\delta\psi_{Bd}=(\gamma^\mu D_\mu Z_d-\frac12 Z_d)\epsilon_{1B}=0\ .
\ee%
In order to have half-BPS configuration the above should vanish
identically for any $\epsilon_{1B}$ (the other components of
$\epsilon$ are not constrained). This is only true if $D_\theta Z =
D_\phi Z = 0$ and $D_0 Z = -(i/2)\,Z$. We note, however, just
checking the $\delta\psi=0$ condition is not enough and one should
make sure that all the equations of motion are also satisfied. For
the $Z$ field this is trivial, but not for the gauge fields (note
that in our \cs theory the gauge fields are non-propagating). In
particular, on top of \eqref{SUSY-variation}, \eqref{F-J-eom} should
also be added, yielding to $F^{(i)}_{t\theta}=F^{(i)}_{t\phi}=0$ and
$F^{(1)}_{\theta\phi} =\frac{\pi }{k}\sin\theta {\bar Z}Z,\ \
F^{(2)}=\frac{\pi }{k}\sin\theta {Z} {{\bar Z}}$. These equations
may be solved in the $A^{(1)}_\mu=A^{(2)}_\mu$ gauge and in the
gauge where $Z$ is diagonal; as was done in the previous section.

After discussing the supersymmetry condition let us also discuss the
classification half-BPS states by the relevant superalgebras. The
\Ne3d theory is invariant under the $3d$ superconformal $Osp(4^*|8)$
superalgebra, which has 32 real supercharges and is the
super-isometry of $AdS_4\times S^7$ geometry. This algebra has a
number of sub-algebras with 16 supercharges which has been listed in
\cite{half-BPS-superalgebras}. The ABJM theory for generic $k$, on
the other hand is an ${\cal N}=6$ superconformal theory and its
superalgebra is $Osp(4^*|6)$, which has 24 real supercharges.

The half-BPS sector we have been studying, which involves only one
of the four complex scalars of the theory, is invariant under
$SU(2|4)$ for the ${\cal N}=8$ case (related to $k=1,2$ ABJM
theories) and $SU(2|3)$ for the ${\cal N}=6$ (generic $k$ ABJM
theory). That is, they fall into singlet representations of the
above-mentioned half-BPS superalgebras. In either cases, the $SU(2)$
invariance is enforced in our construction by demanding invariance
under the $SO(3)$ isometries of the $S^2$ in the radial
quantization. The $SU(4)$ or $SU(3)$ invariance is made manifest in
exciting only one of the four complex scalars. The generator of the
$U(1)$ in these superalgebras is related to $\Delta-J/2$ in the CFT
side and hence its invariance is enforced by imposing the BPS
condition.

\subsection{Construction of half-BPS states}%

In the previous section, we discussed the classical moduli space of
half-BPS configurations consistent with the appropriate bosonic
symmetries. We will now construct gauge invariant operators carrying
the right charges corresponding to these classical configurations.
By the operator--state correspondence, these will describe the
half-BPS states in ABJM theories.

Such operators can only involve a single scalar matrix $Z$. Since
this transforms in the bi-fundamental representation of the
$U(N)\times U(N)$ gauge group, the trace over its matrix indices
will not generate a gauge invariant operator. As already mentioned
in \cite{ABJM}, we can construct local gauge invariant operators
using monopole or 't Hooft operators \cite{extra}. It is this part
of the operator that will carry the information about the magnetic
fluxes turned on in the classical configurations.

Before moving to explicit construction of the relevant monopole
operators to our Chern-Simons theory, we note that turning on fluxes
of the gauge fields, $n_i$ will generically break the \un2\ gauge
symmetry to $u(1)^N\times u(1)^N$. Let us denote the generators of
this remaining Abelian subgroup by $T^1_i$ and $T^2_i$ where
$i=1,2,\cdots, N$. Under $u(1)^N\times u(1)^N$ transformations,
$u(1)^N$ which is generated by $T^1_i+T^2_i$, $Z$ remains invariant
and under those generated by $t_i\equiv T^1_i-T^2_i$, $Z$ rotates by
a phase. In the notations of ABJM $U(1)_b$ is the part of the gauge
symmetry which is generated by $t=\sum_{i=1}^N t_i$. The fluxes
$\{n_i\}$ are then charges of $t_i$. We define the ``total flux"
$J/k$ as
\begin{equation}
  J =k\sum_{i=1}^N n_i\, .
 \label{eq:tflux}
\end{equation}
$J/k$, is hence the flux corresponding to the $U(1)_b$ \cite{ABJM}.

To illustrate the idea behind the construction of these operators,
let us consider the $U(1)\times U(1)$ theory first. Working in the
gauge $A^{(1)}=A^{(2)}=A$ with $A_\theta=0$ and $A_\phi =
n\sin\theta$, the only left gauge transformations are the time
dependent ones acting on $A_0$ as%
 $$ A_0^1\to
A_0^1+\partial_0\lambda, \qquad A_0^2\to
A_0^2+\partial_0\lambda\ .%
$$%
The monopole operator is defined as the imaginary exponential of the
integral of the Chern-Simons form over the 2-sphere and integrating
time from $t\to -\infty$ to a value $t$. Since the gauge field
carries $n$ units of flux, such operator is characterized by
$(k,\,n,\,t)$ :
\be\label{monopole-opt1}%
W_{\text{monopole}}(n;t)\equiv e^{-ik\int_{-\infty}^t dt\ A_0\int
d^2\Omega
F_{\theta\phi}}=e^{ik n\int_{-\infty}^t dt A_0}\ .%
\ee%

From now on, let us focus on the $\cN=8$ theory ($k=1$). Assuming
that all gauge transformations vanish as $t\to -\infty$, i.e.
$\lambda(t\to -\infty)=0$, we conclude the monopole operator
transforms as $W_{monopole}\to e^{-in\lambda} W_{monopole}$ under a
gauge transformation. Recalling that under the same gauge
transformation $Z^n\to e^{+in\lambda}Z^n$, we conclude that
$W(n;t)Z^n(t)$ is a gauge invariant operator.
 This operator has R-charge $n$ and
conformal dimension $n/2$. Notice this is the only gauge invariant
operator carrying these charges one can build for this gauge group
involving a single bi-fundamental matter field. This agrees with the
classical moduli space analysis above. Equivalently, there is a
one--to--one correspondence between the half-BPS operator and the
magnetic flux $n$ that characterizes the charges carried by the
operator.

Let us extend this construction to the $U(N)\times U(N)$ theory,
where we already know the magnetic fluxes are specified by $N$
integers, and not just one as in the Abelian case. In the general
case, turning on the fluxes $\{n_i\}$ generically breaks the gauge
group to $U(1)^N\times U(1)^N$. The individual eigenvalues $z_i$
rotate under the $U(1)$ rotation generated by $t_i$. This suggests
generalizing the above construction for each pair of unbroken
$U(1)\times U(1)$ gauge group factors.

Proceeding as if we have $N$ copies of the $U(1)\times U(1)$ theory
and with a given set of fluxes $\{n_i\}$ the most general gauge
invariant operator involving the monopole operators is hence
\footnote{Note that due to the $SO(3)$ invariance in the half-BPS
sector one may simply reduce the theory on the $S^2$ and remain with
a $0+1$ dimensional quantum mechanical
system. The half-BPS operators are hence operators in this theory and have only
time dependence.}%
 \be\label{BPS-opt-U(N)}%
{\cal O}_{\{n_i\}}=\prod_{i=1}^N W(n_i;t)z_i^{n_i}(t)\ . %
\ee%

To write these operators in a more ``$U(N)\times U(N)$ covariant''
form, let us recall that monopole operators on a $U(N)$ gauge
theory:%
\be\label{O-ti-tf}%
{\cal O}(t_i,t_f)=P\left( e^{i\int_{t_i}^{t_f}\ dt\ A_0}\right),%
\ee%
transform under $U(N)$ gauge transformations as%
\be%
{\cal O}\longrightarrow U(t_i) {\cal O} U(t_f)^{-1}\ .%
\ee%
Thus, if we take $U(-\infty)=\id$, the corresponding operator
\be\label{monopole-operator}%
{\cal O}(t)=P\left( e^{i\int_{-\infty}^{t}\ dt\ A_0}\right)%
\ee%
is in the anti-fundamental of the $U(N)$ gauge group.

Let us now consider our $U(N)\times U(N)$ gauge theory.
 For each gauge group we can construct a 't Hooft operator which is in the
(anti)fundamental of either of the gauge groups. Explicitly consider
\begin{eqnarray}
{\cal O}_1 &=& P\left(e^{(i\int_{-\infty}^t dt A_0^{(1)})}\right)\,, \\
{\cal O}_2 &=& P\left(e^{(-i\int_{-\infty}^t dt
A_0^{(2)})}\right)\,.
\end{eqnarray}
Clearly, ${\cal O}_1$ is in the anti-fundamental of the first $U(N)$
group whereas ${\cal O}_2$ is in the fundamental of the second
$U(N)$. Since $Z$ transforms in the bi-fundamental $(N, {\bar N})$,
we conclude that the operator ${\cal O}_1 Z {\cal O}_2$ is gauge
invariant.

The previous construction has no relation to the individual fluxes
$\{n_i\}$. To introduce the dependence on these quantum numbers, let
us return to the matrix $F^{(a)}$ $a=1,2$. One may use the $U(N)$
large gauge (global) transformations to bring both matrices to a
diagonal form:%
\be%
F^{(a)}|i\rangle_{(a)} = 2\pi n_i^{(a)}\,|i\rangle_{(a)} \ %
\ee%
We can now build projector operators:%
\be%
{\cal P}_i^{(a)}=|i\rangle_{(a)} \langle i|_{(a)}\, .%
\ee%
By construction, each of these projectors transforms in the adjoint
under a gauge transformation of the $a^{th}$ $U(N)$ gauge group.
Thus, the set of states
\begin{equation}
  {\cal Q}_{ij}= {\cal O}_1 P_i^1 Z P_j^2 {\cal O}_2\,,
\end{equation}
is gauge invariant.

To sum up, the product of the traces of these operators raised to
any integer would be a half-BPS gauge invariant operator. However,
due to the equations of motion for the gauge fields
$F^{(1)}=F^{(2)}$. Thus, both basis are equal,
$|i\rangle_{(1)}=|i\rangle_{(2)}$, and we can drop the dependence on
this index. Similarly $Z_{ij}=<i|_{(1)} Z |j>_{(2)}$ is also
diagonal in this same basis.

Denoting by ${\cal Q}={\cal O}_1{\cal O}_2$ the operator
transforming in the $(\bar{\textbf{N}}, \textbf{N})$ representation
of $U(N)\times U(N)$ (using both 't Hooft operators defined above),
we can write our gauge invariant operators as
\be\label{Zi}%
{\cal Z}_i\equiv {\cal Q} {\cal P}_i Z\ .%
\ee%
We can finally write the operators ${\cal O}_{\{n_i\}}$ in terms of the ${\cal Z}_i$ as
\be\label{BPS-opt-U(N)a}%
{\cal O}_{\{n_i\}}=\prod_{i=1}^N {\cal Z}_i^{n_i}\ .
\ee%
Note that $\{n_i\}$ are also gauge invariant quantities.

The set of operators ${\cal O}_{\{n_i\}}$ satisfying $\sum_{i=1}^N
n_i=J$ correspond to half-BPS operators with R-charge $J$. This
explicit construction establishes a one--to--one correspondence
between this class of half-BPS operators and the set of  Young
tableaux with $J$ boxes and at most $N$ rows:  ${\cal O}_{\{n_i\}}$
corresponds to a Young tableau  which has $n_i$ number of boxes in
the $i^{\rm th}$ row. The fact that such Young tableau do completely
characterize the space of vacua of the plane wave Matrix Model
strongly suggests that the operators constructed here are
complete.\footnote{The correspondence to the Matrix Model is known
to be exact in the limit $N\to \infty$, in which the number of rows
(rank of the gauge group) is not fixed.}

For $k\neq 1$ one can check that all the above arguments goes
through and one needs to simply replace $n_i$ in
\eqref{BPS-opt-U(N)a} by $kn_i$.

\section{$2d$ Fermion picture}

The half-BPS sector of $\cN=4$ SYM is described by a matrix model
whose degrees of freedom correspond to free fermions in a one
dimensional harmonic potential or equivalently $2d$ fermions in the
lowest Landau Level. It is natural to wonder whether our $d=3$ SCFTs
have a similar fermionic description for their half-BPS sectors.

Let us assume we can decouple this sector in the full theory and
consider the corresponding partition function. Due to the $SO(3)$
invariance we may do the reduction on the sphere and remain with a
$0+1$ one complex matrix model in the bi-fundamental  of $u(N)\times
u(N)$. As previously discussed, we can use the $U(N)\times U(N)$
gauge symmetry to diagonalize $Z$ and work with its eigenvalues
$z_i$ \eqref{Z-diagonal} as degrees of freedom. The remaining
$U(1)^N$ gauge symmetry can be used to remove the phases of $z_i$
and make them all positive real values, which  will be denoted
by $r_i$. To fix the gauge in which $Z$ is diagonal, however, we
need to include the Jacobian of these gauge transformations into the
measure of the path integral. Being in the bi-fundamental the
procedure is a bit different from the one discussed for adjoint
scalars in $4d$ SYM. This has been carried out in the context of
complex matrix models in \cite{DKKMMS, KMS} and here we sketch the
argument.

To compute the measure factor, we recall the form of
the gauge transformations on $Z$, $Z\to UZ V^{-1}$. Under the
infinitesimal gauge transformations%
\be%
 U\simeq 1+i(\lambda+\rho),\qquad V^{-1}\simeq 1-i(\lambda-\rho),
\ee%
where $\lambda$ and $\rho$ are $N\times N$ hermitian matrices,
elements of $u(N)$ algebra, we have%
\be%
\delta Z=i[\lambda, Z]+i\{\rho, Z\}\ .%
\ee%
Next let us assume that $Z$'s are diagonal:
\be%
Z=diag (z_1,z_2,\cdots, z_N)%
\ee%
and choose the $T^k$ basis for the $N\times N$ gauge
transformations, where the elements of $T^k$ $N\times N$ matrices
are%
\be
(T^k)_{ij}=\delta_{i,j+k},\qquad i+N\equiv i\ .%
\ee%
We then have
\be%
[T^k, Z]=(z_i-z_{i+k})\delta_{i,j+k},\qquad \{T^k,
Z\}=(z_i+z_{i+k})\delta_{i,j+k}\ . %
\ee%
As mentioned earlier, after diagonalization of $Z$ we are still left
with the residual $u(1)^N$ gauge symmetry and one may use that to
bring $z_i$ to  positive real valued $r_i$,
$r_i=|z_i|^2$.\footnote{To be more precise, in the presence of the
Chern-Simons terms we are still left with the over-all $u(1)$, the
$u(1)_b$ which rephases all $z_i$ simultaneously and hence all the
$z_i$ have the same phase.}

The Jacobian of these gauge transformations equals%
\be\label{Jacobian}%
\begin{split}
 J&\equiv |\frac{\delta Z}{\delta \lambda}\cdot
\frac{\delta Z}{\delta \rho}|^2\\ %
&=|\prod_{k=1}^N\prod_{i=1}^k
(z_i-z_{i+k})\cdot (z_i+z_{i+k})|^2=|\prod_{i>j} (z^2_i-z^2_j)|^2\ , %
\end{split}\ee%
The measure of the path integral in the half-BPS sector involves
$DZD\bar Z$ after fixing the gauge and in the basis where $Z$ is
diagonal it becomes $\prod_i dz_i d\bar z_i |\prod_{i>j}
(z^2_i-z^2_j)|^2$. The residual $u(1)^N$ symmetry should now be
implemented. This will not change the Jacobian \eqref{Jacobian} and
its effect is to render $z_i$ real positive and reducing $dz_i d\bar
z_i$ piece to $r_idr_i$. In summary, the gauge fixed measure is
$\prod_i r_i dr_i \prod_{i>j} (r^2_i-r^2_j)^2$ \cite{DKKMMS}.

In analogy with the half-BPS sector of an ${\cal N}=4$ $U(N)$ SYM
theory (e.g. see \cite{LLMvsLLL,Corley:2001zk,Berenstein:2004kk}),
one can then rewrite the partition function of the ABJM theory in the half-BPS sector as%
\be\label{partition-func}%
Z|_{half-BPS}=e^{-F}=\int \left[DA_1DA_2DZD\bar Z\right]_{half-BPS}
e^{-S_{1/2\ BPS}}=\langle
\Psi|\Psi\rangle %
\ee%
where $S_{1/2\ BPS}$ is \eqref{Z-action} but reduced on $S^2$ and
$|\Psi\rangle$ is the wavefunction for the \emph{vacuum state} of a
system of $N$ $2d$ fermions in a harmonic oscillator potential. In
the above by $\left[DA_1DA_2DZD\bar Z\right]_{half-BPS}\equiv D{\cal
M}$ we mean the part of the measure which is allowed by the half-BPS
condition. In other words, we are assuming that the half-BPS sector
is a closed sector of the theory and is protected, even quantum
mechanically, by supersymmetry. In particular, in $D{\cal M}$ we do
not allow for $Z$ and $A$ configurations which have non-vanishing
$D_\theta Z$, $D_\phi Z$, $F^{(i)}_{\theta t}$ or $F^{(i)}_{\phi
t}$. Moreover, $D{\cal M}$ has a
$\delta(2F^{(i)}_{\theta\phi}/\sin\theta-J_0^{(i)})$ factor.
Therefore, what we are computing is effectively the partition
function of a $0+1$ dimensional one complex matrix model which is
exactly equal to the partition function of $N$ $2d$ fermions in a
harmonic oscillator potential. The residual $u(1)^N$ symmetry,
however, amounts to setting the angular momentum of each of these
oscillators on the $2d$ plane equal to zero \cite{DKKMMS}.
Explicitly, each of these fermions should satisfy the Schrodinger
equation%
 \be\label{Schrodinger}%
-\frac{1}{r_i}\partial_{r_i}(r_i\partial_{r_i}
\Psi_i)+r_i^2\Psi_i=2(2n_i+1)\Psi_i%
\ee%
The frequency of this system, as is seen from
\eqref{Hamiltonian-ni}, is $k/4$. Note that working in the zero
angular momentum sector, the energy (once the zero point energy 2 is
subtracted)  is an \emph{even} multiple of the frequency and hence
on the right-hand-side of \eqref{Schrodinger} we have $4$ times an
integer. The state $|\Psi\rangle$ is then obtained from the Slater
determinant of $\Psi_i$'s, which is leading to the measure factor
times a Gaussian with width one. In this picture the factor
$r_idr_i$ is naturally related to the fact that we are working with
$2d$ fermions.

It is worth noting that despite the similarities there are important
differences with the ${\cal N}=4$ SYM case:
\begin{itemize}
\item
In the ${\cal N}=4$ case, the half-BPS sector can be described
through a system of $2d$ fermions  in the presence of a constant
magnetic field in the Lowest Landau Level.\footnote{This was related
to the fact that in the SYM side half-BPS states are holomorphic
functions of one of the three scalars complex scalars of the theory
\cite{LLMvsLLL}.} In our case, degrees of freedom can be interpreted
as $2d$ fermions in an harmonic potential in states of vanishing
angular momentum. Thus, there is no relation to the quantum Hall system nor the Laughlin
wave function. On the other hand, the zero $2d$ angular momentum
condition can be related to a one dimensional ``half harmonic
oscillator potential'' (restricted to move in the $x>0$ region).
However, the latter will not produce the extra $r_i$ factor in the
measure.
\item In the ABJM theory, there is a non-trivial flux over the
$S^2$ coming from the insertion of the 't Hooft loop operators
\eqref{BPS-opt-U(N)}. This should be contrasted with the $\cN=4$ on
$R\times S^3$ (\emph{cf.} footnote \ref{footnote2}). This is
implemented by performing the path integral around the configuration
with these background fluxes (given in \eqref{flux-z2}). This also
leads to the appearance of $k/4$ as the frequency of the $2d$
harmonic oscillator.
\end{itemize}

\section{Half-BPS states in dual descriptions}

The ABJM theory (at level $k=1,2$) is dual to M-theory on
$AdS_4\times S^7$. We have two different available descriptions for
the latter: the $11d$ supergravity in asymptotically $AdS_4\times
S^7$ backgrounds and the plane-wave matrix model \cite{BMN}. In the
following, we will match the half-BPS operators constructed in
previous sections with the description of these states in these
other two formulations of the same system. This will provide a check
of our operator construction in the $N\to \infty$ limit.

\subsection{Plane-wave matrix theory perspective}

The following discussion is strongly based on the results obtained
in \cite{DSV-1,mark-coarsegraining,DSV-2, MSV}. We review them here
for completeness to establish a precise relation with the half-BPS
operators defined before.

The plane-wave matrix model \cite{BMN} is a $0+1$ dimensional $U(N)$ supersymmetric quantum mechanics involving nine scalars $X^A$ and their fermionic counterparts, all in the
$N\times N$ hermitian representation of the $U(N)$ gauge group. The set of scalars is divided into two groups $X^a,\ a=1,2,\cdots, 6$ and $X^i,\ i=1,2,3$. Physical states lie in representations of $SU(2|4)$ comprised of finite collections of representations of the bosonic subalgebra $SO(6)\times SO(3)\times U(1)_H$.

It is known that this matrix model has a discrete set of classical
half-BPS vacua which are interpreted as fuzzy M2-brane spheres. The
half-BPS condition implies $X^a=0$ and the kinetic terms to vanish
and we hence remain with
\be\label{BMN-matrix-model-Xi}%
H=\frac{R_-}{8}\ \text{Tr}\left(i\epsilon^{ijk}[X^i,X^j]+\frac{\mu}{2R_-}X^k\right)^2%
\ee%
($R_-$ is the arbitrary energy scale of the theory and $\mu/R_-$
is the only dimensionless parameter of this theory. These would be
irrelevant to our discussion of half-BPS states.) Zero energy
configurations are hence solutions to%
\be\label{zero-energy-matrix-model}%
[J_i, J_j]=i\epsilon_{ijk}J_k %
\ee%
where $X^i=\frac{\mu}{R_-}J_i$. Thus, all classical vacua are labeled by $J\times J$ reducible representations of $SU(2)$. The latter are determined by a set of $m$ irreducible representations of size $N_i$ appearing $n_i$ times in the decomposition of the initial reducible representation so that
\begin{equation}
  J = \sum_{i=1}^m n_i N_i\,.
 \label{eq:partition}
\end{equation}

Clearly, the set of all classical vacua is equivalent to the problem
of partition  of an integer $J$ into non-negative integers
\cite{MSV}, or equivalently to the set of 2d Young tableau with $J$
boxes. This is exactly the same set characterizing our proposed
half-BPS operators in the ABJM theories. Since it is known that
these states are exact quantum mechanical vacua, and the size of the
representation corresponds to the units of light-cone momentum
carried by the state in its DLCQ interpretation, we can conclude our
matching goes beyond the classical identification.

The microscopic interpretation in terms of (quantized) spherical
M2-branes and M5-branes is similar to the one encountered in the
half-BPS sector of $\cN=4$ SYM. Spherical M2-branes correspond to
dual giant gravitons whose size is proportional to the size of the
irreducible representation $N_i$; $n_i$ stands for the number of
dual giants having the same size. In terms of the Young tableau
description, we can always order the sizes of the irreducible
representations so that $N_i>N_j$ for $i<j$. In this way, a given
Young tableau has $n_i$ rows of length $N_i$, with the total number
of rows $\sum_i n_i$ being the total number of M2-brane giants.
\begin{figure}[th]
\begin{center}
\includegraphics[scale=.75]{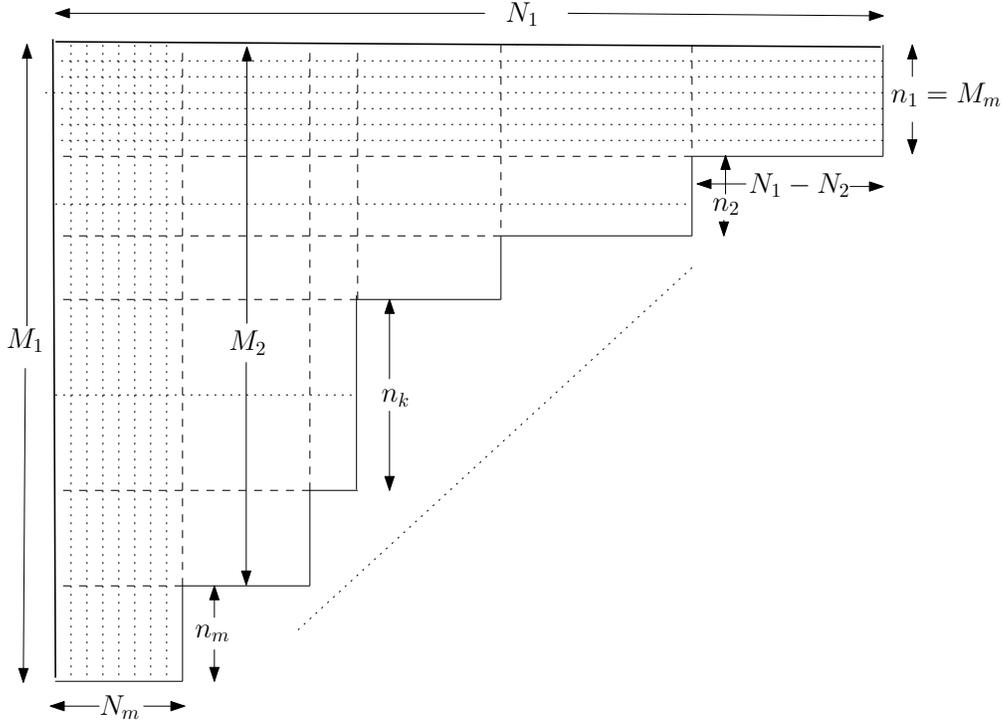}
\caption{A given Young tableau can have interpretation in terms of a
collection of $n_i$ spherical M2-branes of radius $N_i$,
$i=1,2,\cdots, m$ or alternatively in terms of  $m_k$ spherical
M5-branes the radius of which is given by $M_k$, $k=1,2,\cdots, m$.}
\label{M2M5}
\end{center}
\end{figure}

As argued in \cite{MSV}, the same Young tableau and set of labels
can have an interpretation in terms of (quantized) spherical
M5-brane giants, as a collection of $m$ M5-branes the fourth power
of the radius of which is proportional to the amount of the DLCQ
light-cone momentum they carry, $M_k$, and there are $m_k$ five
branes of a given size.\footnote{Note that both size and number of
giants are ``classical'' notions and are not good quantum numbers in
an interacting theory. Even though the M2 or M5 -giant
interpretation is not appropriate one in finite $J$, finite $N_i$ or
finite $M_i$, where the M2 or M5 brane theory becomes strongly
coupled, labeling vacua by these quantum numbers is still meaningful
in such cases.} The M2-brane and M5-brane parameters, as
depicted in Fig.\ref{M2M5}, are related as:%
\be\label{M2M5-radius-size}%
M_k=\sum_{i=1}^{m-k+1} n_i,\qquad m_k=N_{m-k+1}-N_{m-k+2}, \qquad
k=1,2,\cdots, m,\quad N_{m+1}=0.%
\ee%
One can easily check that $\sum_{i=1}^m m_iM_i=\sum_{i=1}^m
n_iN_i=J$. The M5-brane description becomes a good one (weakly
coupled) when the $M_i$ are large, while the M2 description is a
good one when $N_i$ are large.

Modulo the caveats associated with interpreting these states
geometrically as bound states of spherical M2 and M5-brane giants,
we can definitely establish a one--to--one correspondence between
any set of fluxes $\{n_i\}$ determining our half-BPS operators with
the set of dimensions of the irreducible representations
characterizing the plane-wave Matrix model vacua. Note, however,
that in the ABJM theory the corresponding Young tableau has a
maximum number of rows $N$. In the case of the plane-wave matrix
model, as we are dealing with M-theory on the Penrose limit of
$AdS_4\times S^7$, $N$ has been sent to infinity.

\subsection{$11d$ supergravity perspective}

The following discussion is based on \cite{LLM, LM,
mark-coarsegraining}. We include it here for completeness and to
make the emergence of Young tableau from the classical moduli space
of supergravity configurations and its relation to the Young tableau
appearing in our operator construction more explicit. A similar
connection, using the plane-wave matrix model vacua and supergravity
has been discussed in \cite{LM, mark-coarsegraining}.

The classical moduli space of half-BPS configurations preserving
$SO(6)\times SO(3)\times U(1)$ in 11d supergravity was worked out in
\cite{LLM} and we very briefly review them here. These symmetries
reduce the functional dependence of all metric and flux components
to a three dimensional dependence described by a set of coordinates
$\{y,\,x_1,\,x_2\}$. Any solution belonging to this moduli space is
determined by an scalar function $D(y,\,x_1,\,x_2)$ satisfying the
Toda equation
\begin{equation}
(\partial_1^2 + \partial_2^2 )D + \partial_y^2 e^{D} =0\,.
\end{equation}

It was also pointed out that for any translationally or rotationally
invariant configuration, the Toda equation could be mapped through a
non-linear change of variables to a linear 3d Laplace equation. For
our purposes of establishing a dictionary between our proposed
half-BPS operators to supergravity configurations, it will be enough
to restrict ourselves to translationally invariant solutions. These
were extensively studied in  \cite{LM, Mark-M5,mark-coarsegraining}.
In such situation, the Toda equation reduces to
\begin{equation}
  \partial_2^2 D + \partial_y^2 e^{D} = 0\,.
\end{equation}
Using the implicit change of coordinates and
variables
\begin{equation}
  e^{D} = \rho^2\,, \quad \quad y = \rho\partial_\rho V (\rho,\,\eta)\,, \quad \quad x^2 = \partial_\eta V (y,\,\eta)\,,
\end{equation}
the Toda equation is mapped to the axisymmetric 3d Laplace equation
\begin{equation}
  \frac{1}{\rho}\partial_\rho\left(\rho\partial_\rho V\right) + \partial_\eta^2 V = 0\,.
\end{equation}
\begin{figure}[t]
\begin{center}
\includegraphics[scale=.8]{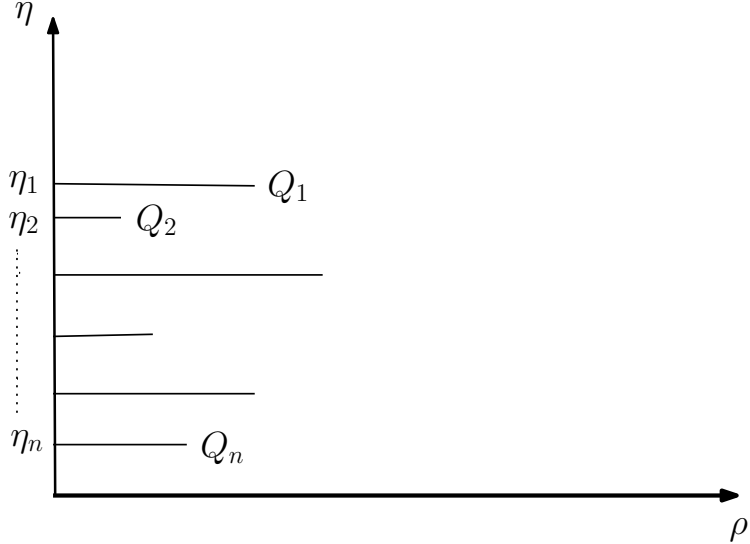}
\caption{A generic half-BPS deformation of the $11d$ plane-wave is
specified with a distribution of conducting disks with charge $Q_i$
located at $\eta_i$  on the $\eta-\rho$ plane.} \label{Fig1}
\end{center}
\end{figure}

This is a problem in electrostatics with potential $V$ which can be
fully determined once we impose a set of boundary conditions that
makes these half-BPS configurations regular. It was shown in
\cite{LM} that this mathematical problem is fixed by specifying the
locations $\eta_i$ and the charges $Q_i$ carried by a discrete set
of conducting disks (their sizes are related to the charges they
carry).

All these solutions will be deformations of the maximally
supersymmetric $11d$ plane-wave background, so let us consider this
solution first. The electrostatic potential is
\be%
V_b=\rho^2\eta-\frac23 \eta^3\ .
\ee%
This fixes the transformation between coordinates to be
\be\label{y-x2-eta-rho}%
y=2\rho^2\eta,\qquad x^2=\rho^2-2\eta^2\ ,%
\ee%
whereas the 11d metric is  
\be\label{plane-wave-eta-rho}%
ds^2= -4(4\eta^2+ \rho^2) dt^2-4 dx^1dt + 4(d\rho^2+d\eta^2)+
4\rho^2 d\Omega_5^2+4\eta^2 d\Omega_2^2\ . %
\ee %
Notice that both $\eta$ and $\rho$ are coordinates related to the
radii of $S^2$ and $S^5$. This point will be important in our
analysis below.

Any excitation on top of this vacuum will be given by a distribution
of conducting disks located at constant $\eta>0$ (see
Fig.\ref{Fig1}).\footnote{Note that the $(\rho,\,\eta)$ plane is
actually a half-plane, since $\rho\geq 0$ and the background
potential $V_b$ fills the  $\eta\leq 0$ region.} The disks locations
are given  by positive $\eta_i$ and their sizes/charges by $\rho_i$
({\emph{cf.} Fig.\ref{Fig1}}). The number of M2-branes $(N_2)$ and
M5-branes $(N_5)$ can be computed in the supergravity approximation
as flux integrals \cite{LM}:
\begin{eqnarray}
N_2 &=& \frac{8Q_i}{\pi^2}\,, \label{M2-brane-number} \\
N_5 &=& \frac{2d_i}{\pi}\,, \label{M5-brane-number}%
\end{eqnarray}
where $Q_i$ is the charge of the $i^\mathrm{th}$ disk and $d_i =
\eta_{i+1}-\eta_i$.
\begin{figure}[t]
\begin{center}
\includegraphics[scale=.75]{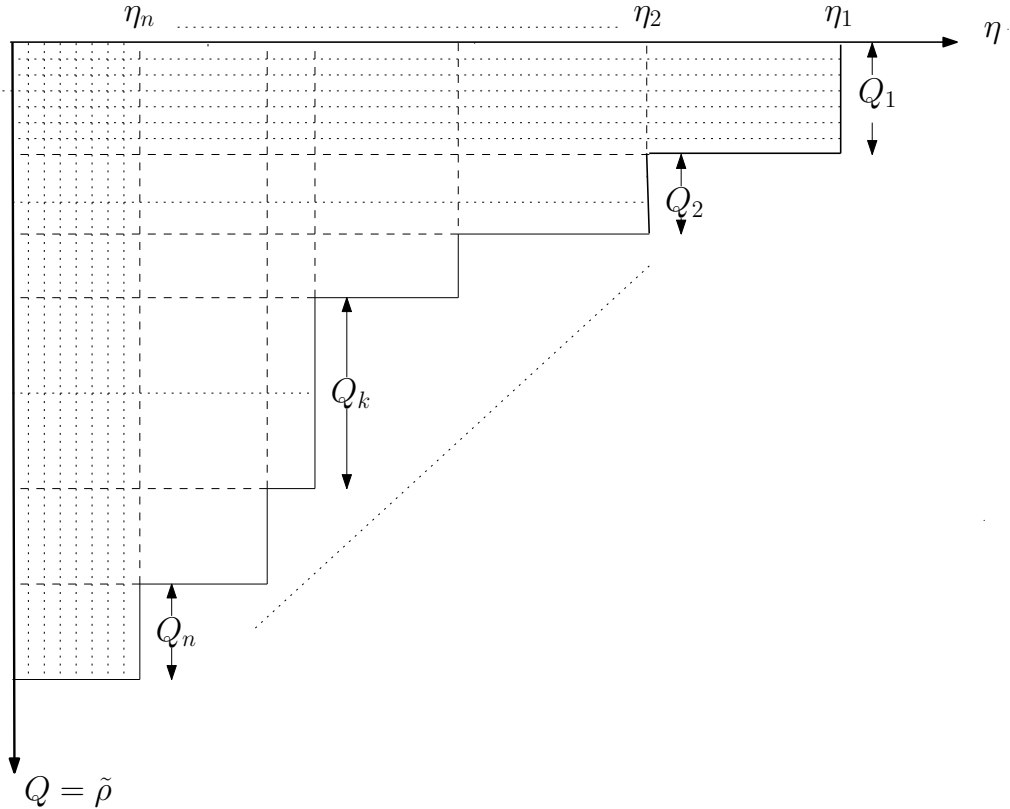}%
\caption{The connection between LLM $\eta-\rho$ plane and the Young
diagram. The length of the rows are determining the size of
M2-branes and the number of row of the same size, $Q_i$, is
determining the number of M2-branes of a given size. Alternatively
one can focus on the columns. The length of columns determine the
size of M5-branes while the number of columns of given length,
$\eta_{i+1}-\eta_i$, determines the number of M5-branes with that
size. The total number of boxes in the Young diagram is hence
$\sum_{i=1}^n \eta_i Q_i$ which is equal to the total R-charge of
the system (above the background value).} \label{Fig2}
\end{center}
\end{figure}
We comment that in the above picture we have considered M2-branes
and M5-branes as perturbations above the background plane-wave and
hence it is meaningful to specify the size and number of branes.

With the above information we can readily map Fig.\ref{Fig1} into
the Young diagram corresponding to the  plane-wave matric theory
vacua. This has been depicted in Fig.\ref{Fig2} (a similar picture
has been discussed in \cite{Mark-M5}). Note that we have chosen the
vertical axis such that it directly measures $Q_i$ rather than the
$\rho$.

\section{Discussion}

We have studied and classified all the $SO(3)$ invariant states in
the half-BPS sector of the $u(N)\times u(N)$ ABJM theory on $R\times
S^2$. First, we determined the classical moduli space of such
configurations and showed that in the sector with R-charge $J$ it
coincides with the solutions  to the problem of partition of $J$
into $N$ non-negative integers. These integers correspond to the
integral fluxes of the gauge fields on the $S^2$ where the theory is
defined. Therefore, these states are in one--to--one correspondence
with Young tableaux of $J$ boxes and maximum number of rows $N$.
Second, we constructed explicit gauge invariant BPS operators
involving  non-Abelian  't Hooft monopole operators.

We showed, through path integral considerations, that partition
function of the ABJM theory in this sector is exactly matching the
partition function of $N$ $2d$ fermions in a harmonic oscillator
potential the frequency of which is $k/4$ and where the fermions are
restricted to move in the zero angular momentum sector. This should
be contrasted with the fermionic picture corresponding to the
half-BPS sector of $\cN=4$ SYM theory. It would be desirable to
match our partition function arguments to the computations of
supersymmetric indices \cite{ABJM-index}  and semi-short operator
counting \cite{Dolan:2008vc} carried for the ABJM theory.

We argued that there is a precise  correspondence between the
half-BPS sectors of ABJM theory, plane-wave matrix theory and the
$11d$ LLM bubbling geometries and all of which can be nicely encoded
in terms of Young diagrams, e.g. see Fig. \ref{Fig2}. This precise
matching was, however, done for $N\to\infty$ case. This was due to
the computational difficulties of constructing the $11d$ LLM
bubbling geometries which are deformations of $AdS_4\times S^7$ (or
$AdS_7\times S^4$), rather than the $11d$ plane-wave. One may use
our results coming from ABJM theory as an additional guide to
construct such solutions. In particular, our analysis of the measure
and the $2d$ fermion picture (for $k=1$ case) suggests that a
similar fermionic picture, as we have in the $10d$ LLM geometries
\cite{LLM}, should also be present in the $11d$ case.

As another related interesting question, for the half-BPS sector in
type IIB on $AdS_5\times S^5$, it was established that the singular
half-BPS superstar supergravity configuration emerges as a
coarse-grained description of the typical state in the Hilbert space
describing $N$ free fermions in the matrix model \cite{library}. In
M-theory, there is a similar singular half-BPS configuration, and it
would be interesting to work out the dictionary between the gravity
data describing the classical moduli space and the classical limit
of the typical quantum states belonging to this sector. The analysis
of such ``typical states'' in the gauge theory side using the
plane-wave matrix model was studied in \cite{mark-coarsegraining}.
If the wave functions for these states do not get renormalized, it
should be possible to establish a connection between gauge theory
and gravity. If there is renormalization, such connection may not be
so apparent as for the $\cN=4$ story described in \cite{library}.

As argued the matrix theory and supergravity descriptions can be
interpreted in terms of M2-branes or M5-branes. In the ABJM theory,
being a $2+1$ dimensional field theory, the more natural
interpretation is in terms of M2-branes. It is interesting to
elaborate further on the M5-brane picture in the ABJM theory. One
specific computation in this direction could be studying the
spectrum of fluctuations of the theory around its half-BPS vacua. A
similar analysis within the plane-wave matrix theory revealed
\cite{MSV} that among these fluctuations those which are protected
by supersymmetry may be identified with the fluctuations of
spherical M5-branes.

As argued in \cite{half-BPS-superalgebras,Dolan:2008vc} the $Osp(4^*|8)$ has
other half-BPS superalgebras than  $SU(2|4)$. It would be
interesting to study the moduli space of half-BPS states which are
invariant under these other half-BPS superalgebras.
 As a direct continuation of our half-BPS analysis one may also study
and classify less BPS states. This problem has been considered e.g.
in \cite{Berenstein:2008dc} and \cite{ABJM-index}.

\section*{Acknowledgements}

We would like to thank Hai Lin for his collaboration at the early
stages of this work and David Berenstein, Juan Maldacena and Ofer
Aharony for comments and discussions. M.M.Sh-J. would like to thank
the Abdus Salam ICTP where a part of this work was carried out. The
work of J.S. was partially supported by the Engineering and Physical
Sciences Research Council [grant number EP/G007985/1]. This research
was supported in part by the National Science Foundation under Grant
No. NSF PHY05-51164. J.S. would like to thank the organizers of the
KITP programme ``Fundamentals of String Theory'' for hospitality
during the final stages of this project.



\begin{thebibliography}{99}

\bibitem{ABJM}
  O.~Aharony, O.~Bergman, D.~L.~Jafferis and J.~Maldacena,
  ``N=6 superconformal Chern-Simons-matter theories, M2-branes and their
  gravity duals,''
  JHEP {\bf 0810}, 091 (2008)
  [arXiv:0806.1218 [hep-th]].

\bibitem{LLM}
  H.~Lin, O.~Lunin and J.~M.~Maldacena,
  ``Bubbling AdS space and half-BPS geometries,''
  JHEP {\bf 0410}, 025 (2004)
  [arXiv:hep-th/0409174].

\bibitem{BLG}
  J.~Bagger and N.~Lambert,
  ``Modeling multiple M2's,''
  Phys.\ Rev.\  D {\bf 75} (2007) 045020
  [arXiv:hep-th/0611108];
  ``Gauge Symmetry and Supersymmetry of Multiple M2-Branes,''
  Phys.\ Rev.\  D {\bf 77} (2008) 065008
  [arXiv:0711.0955 [hep-th]].
  A.~Gustavsson,
  ``Algebraic structures on parallel M2-branes,''
  arXiv:0709.1260 [hep-th];
  ``Selfdual strings and loop space Nahm equations,''
  arXiv:0802.3456 [hep-th].

\bibitem{3algebra-nogo}
J.~P.~Gauntlett and J.~B.~Gutowski,
  ``Constraining Maximally Supersymmetric Membrane Actions,''
  arXiv:0804.3078 [hep-th].

 G.~Papadopoulos,
  ``M2-branes, 3-Lie Algebras and Plucker relations,''
  arXiv:0804.2662 [hep-th].

\bibitem{Mark}
  M.~Van Raamsdonk,
  ``Comments on the Bagger-Lambert theory and multiple M2-branes,''
JHEP {\bf 0805}, 105 (2008),  arXiv:0803.3803 [hep-th].

\bibitem{Schwarz-SUSY}
  M.~A.~Bandres, A.~E.~Lipstein and J.~H.~Schwarz,
  ``Studies of the ABJM Theory in a Formulation with Manifest SU(4)
  R-Symmetry,''
  JHEP {\bf 0809}, 027 (2008)
  [arXiv:0807.0880 [hep-th]].

\bibitem{BMN}
  D.~E.~Berenstein, J.~M.~Maldacena and H.~S.~Nastase,
  ``Strings in flat space and pp waves from N = 4 super Yang Mills,''
  JHEP {\bf 0204}, 013 (2002)
  [arXiv:hep-th/0202021].

\bibitem{extra}
  V.~Borokhov, A.~Kapustin and X.~k.~Wu,
  ``Topological disorder operators in three-dimensional conformal field
  theory,''
  JHEP {\bf 0211} (2002) 049
  [arXiv:hep-th/0206054];
  ``Monopole operators and mirror symmetry in three dimensions,''
  JHEP {\bf 0212} (2002) 044
  [arXiv:hep-th/0207074].

  N.~Itzhaki,
  ``Anyons, 't Hooft loops and a generalized connection in three dimensions,''
  Phys.\ Rev.\  D {\bf 67} (2003) 065008
  [arXiv:hep-th/0211140].

\bibitem{recent}
Y.~Imamura, ``Monopole operators in $\cN=4$ Chern-Simons theories
and wrapped M2-branes,''
  arXiv:0902.4173 [hep-th];
    ``Charges and homologies in $AdS_4/CFT_3$,''
  arXiv:0903.3095 [hep-th].



\bibitem{LLMvsLLL}
  A.~Ghodsi, A.~E.~Mosaffa, O.~Saremi and M.~M.~Sheikh-Jabbari,
``LLL vs. LLM: half-BPS sector of N = 4 SYM equals to quantum Hall
system,''
  Nucl.\ Phys.\  B {\bf 729}, 467 (2005)
  [arXiv:hep-th/0505129].

\bibitem{TGMT}
A.~Shomer,
  ``Penrose limit and DLCQ of string theory,''
  Phys.\ Rev.\  D {\bf 68}, 086002 (2003)
  [arXiv:hep-th/0303055].

  M.~M.~Sheikh-Jabbari,
  ``Tiny graviton matrix theory: DLCQ of IIB plane-wave string theory, a
  conjecture,''
  JHEP {\bf 0409}, 017 (2004)
  [arXiv:hep-th/0406214].


\bibitem{DSV-1}
  K.~Dasgupta, M.~M.~Sheikh-Jabbari and M.~Van Raamsdonk,
  ``Matrix perturbation theory for M-theory on a PP-wave,''
  JHEP {\bf 0205}, 056 (2002)
  [arXiv:hep-th/0205185].

\bibitem{LM}
  H.~Lin and J.~M.~Maldacena,
  ``Fivebranes from gauge theory,''
  Phys.\ Rev.\  D {\bf 74}, 084014 (2006)
  [arXiv:hep-th/0509235].

\bibitem{Mark-M5}
  H.~Ling, A.~R.~Mohazab, H.~H.~Shieh, G.~van Anders and M.~Van Raamsdonk,
  ``Little string theory from a double-scaled matrix model,''
  JHEP {\bf 0610}, 018 (2006)
  [arXiv:hep-th/0606014].

\bibitem{mark-coarsegraining}
H.~H.~Shieh, G.~van Anders and M.~Van Raamsdonk,
 ``Coarse-Graining the Lin-Maldacena Geometries,''
  JHEP {\bf 0709}, 059 (2007)
  [arXiv:0705.4308 [hep-th]].

\bibitem{half-BPS-superalgebras}
  E.~D'Hoker, J.~Estes, M.~Gutperle, D.~Krym and P.~Sorba,
  ``Half-BPS supergravity solutions and superalgebras,''
  JHEP {\bf 0812}, 047 (2008)
  [arXiv:0810.1484 [hep-th]].

\bibitem{Ahmad}
 M.~Ali-Akbari,
 ``3d CFT and Multi M2-brane Theory on $R\times S^2$,''
  arXiv:0902.2869 [hep-th].


\bibitem{Berenstein:2008dc}
  D.~Berenstein and D.~Trancanelli,
  ``Three-dimensional N=6 SCFT's and their membrane dynamics,''
  arXiv:0808.2503 [hep-th].

\bibitem{BL3}%
  J.~Bagger and N.~Lambert,
  ``Three-Algebras and N=6 Chern-Simons Gauge Theories,''
  Phys.\ Rev.\  D {\bf 79}, 025002 (2009)
  [arXiv:0807.0163 [hep-th]].

 M.~M.~Sheikh-Jabbari,
  ``A New Three-Algebra Representation for the ${\cal N}=6\ ,
  su(N)\times su(N)$
  Superconformal Chern-Simons Theory,''
  JHEP {\bf 0812}, 111 (2008)
  [arXiv:0810.3782 [hep-th]].

\bibitem{DKKMMS}
M.~R.~Douglas, I.~R.~Klebanov, D.~Kutasov, J.~M.~Maldacena,
E.~J.~Martinec and N.~Seiberg,
  ``A new hat for the c = 1 matrix model,''
  arXiv:hep-th/0307195.

\bibitem{KMS}
  I.~R.~Klebanov, J.~M.~Maldacena and N.~Seiberg,
  ``Unitary and complex matrix models as 1-d type 0 strings,''
  Commun.\ Math.\ Phys.\  {\bf 252}, 275 (2004)
  [arXiv:hep-th/0309168].

\bibitem{Corley:2001zk}
  S.~Corley, A.~Jevicki and S.~Ramgoolam,
  ``Exact correlators of giant gravitons from dual N = 4 SYM theory,''
  Adv.\ Theor.\ Math.\ Phys.\  {\bf 5} (2002) 809
  [arXiv:hep-th/0111222].

\bibitem{Berenstein:2004kk}
  D.~Berenstein,
  ``A toy model for the AdS/CFT correspondence,''
  JHEP {\bf 0407} (2004) 018
  [arXiv:hep-th/0403110].


\bibitem{DSV-2}
  K.~Dasgupta, M.~M.~Sheikh-Jabbari and M.~Van Raamsdonk,
  ``Protected multiplets of M-theory on a plane wave,''
  JHEP {\bf 0209}, 021 (2002)
  [arXiv:hep-th/0207050].

\bibitem{MSV}
  J.~M.~Maldacena, M.~M.~Sheikh-Jabbari and M.~Van Raamsdonk,
  ``Transverse fivebranes in matrix theory,''
  JHEP {\bf 0301}, 038 (2003)
  [arXiv:hep-th/0211139].

\bibitem{Jaume-Gomis}
N.~Drukker, J.~Gomis and D.~Young,
  ``Vortex Loop Operators, M2-branes and Holography,''
  arXiv:0810.4344 [hep-th].

\bibitem{ABJM-index}
  J.~Bhattacharya and S.~Minwalla,
  ``Superconformal Indices for ${\cal N}=6$ Chern Simons Theories,''
  JHEP {\bf 0901}, 014 (2009)
  [arXiv:0806.3251 [hep-th]].

S.~Kim,
  ``The complete superconformal index for $\cN=6$ Chern-Simons theory,''
  arXiv:0903.4172 [hep-th].

\bibitem{Dolan:2008vc}
  F.~A.~Dolan,
  ``On Superconformal Characters and Partition Functions in Three Dimensions,''
  arXiv:0811.2740 [hep-th].

\bibitem{library}
  V.~Balasubramanian, J.~de Boer, V.~Jejjala and J.~Simon,
  ``The library of Babel: On the origin of gravitational thermodynamics,''
  JHEP {\bf 0512}, 006 (2005)
  [arXiv:hep-th/0508023].


\end{thebibliography}
\end{document}